\documentclass[12pt]{article}
\usepackage{amssymb}
\usepackage{amsmath}
\usepackage{cite}
\usepackage{epsfig}
\makeatletter
\def\@cite#1#2{{[{#1}]\if@tempswa\typeout
{IJCGA warning: optional citation argument
ignored: `#2'} \fi}}

\newcount\@tempcntc
\def\@citex[#1]#2{\if@filesw\immediate\write\@auxout{\string\citation{#2}}\fi
  \@tempcnta\z@\@tempcntb\m@ne\def\@citea{}\@cite{\@for\@citeb:=#2\do
    {\@ifundefined
       {b@\@citeb}{\@citeo\@tempcntb\m@ne\@citea\def\@citea{,}{\bf ?}\@warning
       {Citation `\@citeb' on page \thepage \space undefined}}%
    {\setbox\z@\hbox{\global\@tempcntc0\csname b@\@citeb\endcsname\relax}%
     \ifnum\@tempcntc=\z@ \@citeo\@tempcntb\m@ne
       \@citea\def\@citea{,}\hbox{\csname b@\@citeb\endcsname}%
     \else
      \advance\@tempcntb\@ne
      \ifnum\@tempcntb=\@tempcntc
      \else\advance\@tempcntb\m@ne\@citeo
      \@tempcnta\@tempcntc\@tempcntb\@tempcntc\fi\fi}}\@citeo}{#1}}
\def\@citeo{\ifnum\@tempcnta>\@tempcntb\else\@citea\def\@citea{,}%
  \ifnum\@tempcnta=\@tempcntb\the\@tempcnta\else
   {\advance\@tempcnta\@ne\ifnum\@tempcnta=\@tempcntb \else
\def\@citea{--}\fi
    \advance\@tempcnta\m@ne\the\@tempcnta\@citea\the\@tempcntb}\fi\fi}
\makeatother

\newcommand{\gsim}{\lower.7ex\hbox{$\;\stackrel{\textstyle>}{\sim}\;$}}
\newcommand{\lsim}{\lower.7ex\hbox{$\;\stackrel{\textstyle<}{\sim}\;$}}
\newcommand{\be}{\begin{equation}}
\newcommand{\ee}{\end{equation}}
\newcommand{\bea}{\begin{eqnarray}}
\newcommand{\eea}{\end{eqnarray}}

\usepackage{amssymb}

\def\baselinestretch{1}
\begin{document}
\catcode`@=11
\newtoks\@stequation
\def\subequations{\refstepcounter{equation}%
\edef\@savedequation{\the\c@equation}%
  \@stequation=\expandafter{\theequation}
  \edef\@savedtheequation{\the\@stequation}
  \edef\oldtheequation{\theequation}%
  \setcounter{equation}{0}%
  \def\theequation{\oldtheequation\alph{equation}}}
\def\endsubequations{\setcounter{equation}{\@savedequation}%
  \@stequation=\expandafter{\@savedtheequation}%
  \edef\theequation{\the\@stequation}\global\@ignoretrue

\noindent}
\catcode`@=12
\begin{titlepage}
\title{{\bf Gravity on codimension 2 brane worlds}}
\vskip2in
\author{
\centerline{
{\bf Ignacio Navarro$^a$\footnote{\baselineskip=16pt E-mail: {\tt
ignacio.navarro@durham.ac.uk}}}
$\;\;$and$\;\;$
{\bf Jos\'e Santiago$^{a,b}$\footnote{\baselineskip=16pt E-mail: {\tt
jsantiag@fnal.gov}}}
}
\vspace{.35cm}
\\
 $^a${\small IPPP, University of Durham, DH1 3LE Durham, UK}.\\
$^b${\small Fermi National Accelerator Laboratory, P.O. Box 500,
  Batavia, IL 60510, USA}. 
}

\date{}
\maketitle
\def\baselinestretch{1.15}
\begin{abstract}
\noindent We compute the matching
conditions for a general thick codimension 2 brane, 
a necessary previous step towards the investigation of gravitational
phenomena in codimension 2 braneworlds.  
We show that, provided the brane is weakly curved, they are specified 
by the integral in the extra dimensions of the brane energy-momentum,
independently of its detailed 
internal structure. These general matching conditions can then be used
as boundary 
conditions for the bulk solution. By evaluating 
Einstein equations at the brane boundary we are able to write an
evolution equation for the induced metric on the brane depending only
on physical brane parameters and the bulk energy-momentum tensor. We
particularise to a cosmological metric and 
show that a realistic cosmology can be obtained in the simplest
case of having just a non-zero cosmological constant in the bulk.
We point out
several parallelisms between this case and the codimension 1 brane
worlds in an AdS space.

\end{abstract}

\thispagestyle{empty} \vspace{5cm}  \leftline{}

\vskip-22cm \rightline{} \rightline{IPPP/04/48,
DCPT/04/96} 
\rightline{FERMILAB-PUB-04-350-T} 
\rightline{hep-th/0411250} \vskip3in

\end{titlepage}
\setcounter{footnote}{0} \setcounter{page}{1}
\newpage
\baselineskip=20pt

\section{Introduction}

New cosmological observations seem to imply that the expansion of our universe
is currently accelerating \cite{Spergel:2003cb}, driven by a dominant
component of the 
energy-momentum tensor (EMT) with an equation-of-state parameter $w$
close to -1 (the so called dark energy). The observations have made
the cosmological 
constant problem a very pressing one: to the traditional puzzle of an
(almost) vanishing vacuum energy now cosmologists (and particle
physicists) wonder why is its magnitude comparable to the matter
energy density $today$. Recent analysis of the data
\cite{Hannestad:2004cb} point to an even more bizarre situation: the
best fit to observations agrees with a dark energy equation of state
with $\omega<-1$. So the family of problems associated with the vacuum
energy (the cosmological constant problem and its smaller cousin, the
coincidence problem) could grow in the near future with a new member:
why is the vacuum energy $growing$ in time? The problem with this
possibility is that it is not easily accommodated in generally
covariant theories 
as long as the matter EMT satisfies the usual energy conditions
\cite{Carroll:2004hc}. In 
the same fashion, other existing observations also suggest modified
gravitational dynamics\footnote{One can mention the flatness of galaxy
  rotation curves, that can be explained using the dark matter
  hypothesis but that can also be regarded as pointing towards modified
  gravitational dynamics \cite{Scarpa:2003yw}, or the more nearby
  measured Pioneer 10/11 
  anomalies \cite{Anderson:1998jd}.}.  

Weinberg's theorem \cite{Weinberg:1988cp} shows that standard
approaches (by which we mean 4D field theories based on
  General Relativity) to the cosmological
constant problem are very likely to fail and therefore more exotic
ones should be tried. In particular having more than four
dimensions in a Kaluza-Klein fashion does not seem to improve the
situation, since the extra dimensions are small and gravity is
effectively four dimensional below some scale. In this effective
theory one will face the same problems as in any 4D theory. Thus, within
this class of theories, anthropic considerations seem at the moment
the only framework capable of explaining some of the large scale 
properties of our universe \cite{Weinberg:1987dv}.
Brane-world gravity, on the other hand, does not belong to this class
of theories, since it is not guaranteed that the low energy
description of gravity can be obtained from a generally covariant
four-dimensional Lagrangian, $i.e.$, there is \textit{not} necessarily 
a four dimensional description of the gravitational sector. In these
models one assumes that the 
Standard Model fields are confined to some submanifold of the whole
space-time. One can think on the Standard Model fields as the zero
modes of topological defects of higher dimensional field theories
\cite{Rubakov:1983bb} or the gauge theories living on the world-volume
of the string theory D-branes \cite{Polchinski:1998rr}. Fermionic
fields and gauge interactions can be in this way clearly lower
dimensional but to elucidate if 
the gravitational interactions can be well
approximated by 4D Einstein gravity is not so obvious. The hope is
to find in this context
a theory that
shares the good conceptual advantages of 4D General Relativity
(gravity as geometry, background independence...) but can yield a
realistic but non-standard cosmology (or gravitational dynamics).

The cosmology
of codimension one 
braneworlds is quite well understood. It is possible to recover
something close to standard cosmology with  
corrections that take the form of ``dark radiation'' plus terms
involving the matter energy-momentum tensor squared
\cite{Shiromizu:1999wj,Binetruy:1999ut}. Although the situation is not
as good in the 
understanding of codimension two brane worlds, 
great progress has been made recently in
their investigation. The fact that
one can find solutions for a flat brane in a given setup for any value
of its tension has encouraged many authors to try and attack the
cosmological constant problem in
such scenario
\cite{Chen:2000at,Gibbons:2003di,Cline:2003ak,Bostock:2003cv,Vinet:2004bk,Kanno:2004nr}. 
The effect of the brane tension in these models is simply to produce a
deficit angle in the transverse space, without further implications
for its induced metric. The two dimensional space transverse to the
brane acquires locally the geometry of a cone, with the brane situated
on its tip. However one problem of codimension two brane-worlds is that
with a deficit angle one can only generate a two dimensional 
delta function in the Einstein tensor that is proportional to the
brane induced metric. This means that one can only find nonsingular
solutions if the brane EMT is proportional to its
induced metric, $i.e.$ it is pure tension
\cite{Cline:2003ak,Bostock:2003cv}. Thus the solutions found in
\cite{Chen:2000at} cannot be extended to a general brane
EMT in the thin brane limit.
This limit is indeed
singular for a general brane and, as such,
makes all the arguments about the nature of gravity (and self-tuning)
in codimension two braneworlds in Einstein gravity dodgy when working
with $\delta$-like branes. To make things worse, the very Einstein
equations imply that in the 
case of an infinitesimally thin pure tension brane, the deficit angle
is space and time independent. This situation is very similar to the
cosmic string models studied in 4D (see \cite{Geroch:1987qn} for a
rigorous treatment of codimension one and two sources in 4D General
Relativity). It is therefore not sensible to ask 
questions like what happens if there is a sudden phase transition that
changes the tension of the brane as such thing is not allowed by the
equations of motion in the thin brane limit. In other words,
self-tuning has to be formulated as a dynamical process and delta-like
codimension two branes do not allow to do that.
In \cite{Bostock:2003cv} a possible way out
of this situation was proposed by 
adding the Gauss-Bonnet term to the Einstein-Hilbert Lagrangian. In this
case, the thin brane limit is well defined and one finds the
remarkable result that four dimensional Einstein
gravity is recovered as the dynamics for the induced
metric on the infinitesimally thin brane. (Similar conclusions have
been obtained in~\cite{Corradini:2001qv} at the linearised level.)
See also \cite{Kim:2001rm,Navarro:2004kh} for a different approach to
codimension 2 braneworlds in Gauss-Bonnet gravity.

Motivated by the previous considerations, we abandon in this paper 
the thin brane idealization, and compute the
``matching'' conditions for a general thick codimension two brane in Einstein
gravity\footnote{We do not introduce the Gauss-Bonnet term now because
  once one considers a thick brane the Gauss-Bonnet contributions,
  although crucial in the thin limit for obtaining a regular geometry,
  will 
  be subleading unless the Gauss-Bonnet coupling is very large or the
  brane is extremely thin.}. By matching conditions we mean equations
that relate the values of the first derivatives of the metric (with
respect to the orthogonal coordinate, $r$) at the brane boundary with
the brane EMT. In fact, since we are dealing with a thick defect (with
a singular thin limit), 
it is not clear that one should be able to do this without knowing all the
precise small scale structure of the brane. If we need this
information in order to find the solution, the 
matching condition approach would be of no use, since one should solve
the equations independently for different microscopic brane models,
and one could  obtain different results for different models even if
the total energy-momentum carried by the brane is the
same. Gravitational physics would then in general depend on the
ultraviolet details of our theory and therefore model independent
assertions would be difficult to make.
We will
see, however, that one can obtain this set of equations depending only
on the integral of the brane EMT along the extra dimensions as long as
the parallel derivatives (\textit{i.e.} with respect to the brane
coordinates) of the metric, and in particular the Ricci tensor of the 
brane induced metric,
are small enough: in this case our matching conditions do not depend on
the inner structure of the brane. How small is ``small enough'' will
be made clear in the next section but one can argue that this
situation is quite general in the sense that 
it is natural that the presence of the
brane induces much larger gradients in the radial (transverse) 
direction that in the longitudinal ones. 
In
particular this is clearly the case if we are interested in late time
cosmology. Of course that does not mean that nothing 
can be said about very early cosmological times or other situations with
strong gravity effects, but one should keep in mind when dealing with
such situations that our matching conditions imply certain
approximations that break down when the 4D curvature is very large. 
Once we consider a thick brane, departure from pure tension is
allowed and the 
deficit angle can develop a space-time dependence, thus 
questions about the self-tuning behaviour of the system are again legitimate.

In the next section 
we will carefully explain our assumptions and approximations and we
will obtain the equations that relate the $total$ brane
energy-momentum with the deformation of spacetime it causes, the so
called matching 
conditions. We will 
specialize these equations to the cosmological case in the third
section. Evaluating Einstein equations just outside the brane and
using the matching conditions we will be able to obtain the equations
that govern the cosmological evolution of this braneworld. In this
respect, we follow a procedure completely analogous to that of the
codimension one case~\cite{Binetruy:1999ut}.

\section{Matching conditions for a codimension 2 brane}

In this section we will try to answer the following questions: what is
the effect on space-time of an energy-momentum distribution that can
be interpreted as a codimension 2 defect in six-dimensional Einstein
gravity? Is there a way to characterise this effect without knowing
all the precise small-scale structure of the brane? We will see that
the answer to the second question is yes, provided certain conditions
are met, and we will also provide a 
(partial) 
answer to the first question. The
needed conditions have the interpretation of requiring
a weakly curved brane.

In this section we will consider the following quite general ansatz
for the metric,
\begin{equation}
d s^2=g_{\mu\nu}(x,r) dx^\mu dx^\nu-dr^2-L(x,r)^2 d\theta^2,
\end{equation}
where, as usual, $x^\mu$ denote four non-compact dimensions (including a
time-like one), $\mu=0,\ldots,3$, whereas $r,\, \theta$ denote the radial
and angular coordinates of the (compact or not) two extra
dimensions. This means that in particular the following boundary
conditions hold (in order to avoid singularities) at $r=0$ 
\begin{equation}
\quad L(x,0)=0,\quad L^\prime(x,0)=1, \quad \partial_r g_{\mu\nu}(x,0)=0,
\label{boundary:cond:origin}
\end{equation}
where a prime denotes derivative with respect to $r$. We have assumed
a rotational symmetry around the codimension 2 submanifold defined
by the condition $r=0$. 
The metric is determined by the Einstein equation, 
that can be written as
\begin{equation}
M_\ast^4 R^M_N=T^M_N -\frac{1}{4}\delta^M_N T,
\end{equation}
where $M_\ast$ is the 6D fundamental mass, $T_{MN}$ is the EMT and
$T\equiv T^{M}_{\; M}$ 
its trace. 
The brane will be a cylindrically symmetric extended object that
fills the region with $r<\epsilon$.
Since we are trying to deal here with a general situation, and we do
not have a particular microscopic theory for the brane, we cannot
provide a precise definition of the brane width parameter, $\epsilon$,
but in a given particular model it should not be too hard to provide a
strict definition of it. In any case, the same results should be
obtained taking a different definition of the brane width ($i.e$ a
different splitting into brane-bulk of the total 6D EMT), as long as
we are considering the same energy-momentum distribution. 

The presence
of the brane 
induces a strong $r$-dependence of the
curvature tensor that should be reflected on large $r$-derivatives of
the metric. It is natural then to assume that the brane has the effect
of producing mainly non-zero 
$r$-derivatives of the metric, and these derivatives are the relevant
terms in the Einstein equations when looking for a solution. So, given
the boundary conditions at $r=0$, 
eq.(\ref{boundary:cond:origin}), we would like to obtain the values
for the first derivatives of the metric at $r=\epsilon$ in terms of
the brane EMT~\footnote{We call brane EMT to $T_{MN}(r < \epsilon)$,
  including possible contributions from the bulk EMT inside the
  extension of the brane.}. 
For
doing this we follow the standard procedure of integrating the
equations of motion in the region with $r<\epsilon$. Notice that if we
are to find a result that is independent of the inner structure of the
brane, the dominant terms of the integral should be total
$r$-derivatives. If this is the case the value of the integral depends
only on the value of some functions on the boundary, $r=\epsilon$ (and
the origin, $r=0$), and not on the precise solution inside the brane.  

The set of equations we will have to deal with, for the metric at
question, will be given next. We will offer this equations in a form
that makes transparent which terms can be integrated exactly and which
ones should be neglected when dealing with the matching. The $\mu\nu$
components of  
the Ricci
tensor can be written as,
\begin{equation}
\sqrt{g} L R^\mu_\nu = \frac{1}{2} [ \sqrt{g} L K^\mu_\nu]^\prime
+\sqrt{g}L R^\mu_\nu(g)-\sqrt{g}\nabla^\mu \nabla_\nu L,
\label{Gmunu}
\end{equation}
where $K_{\mu\nu}\equiv \partial_r g_{\mu\nu}$ (we will also use
$K\equiv K^\mu_\mu$),
$\nabla_\mu$ denote 
four-dimensional (\textit{i.e.} 
with respect to the metric $g_{\mu\nu}$) covariant
derivatives and $R_{\mu\nu}(g)$ is the Ricci tensor for the
four-dimensional metric $g_{\mu\nu}$.
The $\theta\theta$ component of Ricci tensor reads
\begin{equation}
\sqrt{g}LR^\theta_\theta = [\sqrt{g}L^\prime]^\prime
-\sqrt{g} \nabla^\rho \nabla_\rho L.
\label{Gthetatheta}
\end{equation}
We see that these two equations can be written as a total $r$
derivative plus terms involving only derivatives with respect to the
longitudinal coordinates ($x^\mu$). As we will see in a moment, they
will determine the matching conditions. As for the other two
non-vanishing components of Einstein equations for our metric, 
the $rr$ and $\mu r$ components of the Ricci tensor read,
respectively,
\begin{equation}
R^r_r=
\frac{L^{\prime\prime}}{L}+\frac{1}{2} K^\prime+\frac{1}{4}
K^\rho_\sigma K^\sigma_\rho, \label{Rrr}
\end{equation}
and
\begin{equation}
R_{\mu r}=-  \frac{\partial_\mu L^\prime}{L} + \frac{1}{2}
\frac{\partial_\nu L}{L} K^\nu_\mu + \frac{1}{2} \nabla^\nu
(K_{\mu\nu} -g_{\mu\nu} K).\label{Gtr} 
\end{equation}

We can now integrate the $\mu\nu$ and $\theta\theta$ components of the
Ricci tensor in the region 
$0 \leq r \leq \epsilon$ (the integration in $\theta$ is trivial) 
to find the desired matching conditions.
We start with the $\mu\nu$ components, eq.(\ref{Gmunu}). Integrating
this equation and neglecting the terms that do not have
$r$-derivatives\footnote{For the cosmological case we will estimate
  $R_{\nu}^{\mu}(g)$ and the other neglected terms of 
  eqs.(\ref{Gmunu},\ref{Gthetatheta}) that involve $\mu$-derivatives,
  but to obtain the 
  cosmological evolution we need first the matching conditions. We
  will see at the end that these terms are indeed negligible (with
  respect to the terms we have kept) in the 
  cosmological solutions we consider. This is not surprising since once
  one imposes the constraints of having a realistic late-time
  cosmology the 4D curvature or $\mu$-dependence of the solution have
  to be extremely small.} one gets 
\begin{equation}
\frac{2\pi}{\sqrt{g}|_\epsilon}\int_0^\epsilon dr\, \sqrt{g} L R^\mu_\nu 
\simeq \pi \frac{K^\mu_\nu|_\epsilon}{M_b} 
\simeq \frac{1}{M_\ast^4\sqrt{g}|_\epsilon} 
\int_0^{2\pi} d\theta
\int_0^\epsilon dr\, \sqrt{g}L \left[ T^\mu_\nu -\frac{1}{4} \delta^\mu_\nu
  T \right]
\equiv 
\frac{\hat{T}^\mu_\nu-\frac{1}{4} \delta^\mu_\nu \hat{T}}{M_\ast^4},
\label{matching:munu}
\end{equation}
where we have defined $L(x,\epsilon)\equiv 1/M_b$ (notice that $1/M_b
\sim \epsilon$) and it is understood in here
and in the following that the subscript $|_p$ means that the 
corresponding function is evaluated at $r=p$. We have also
defined the 4D brane EMT, $\hat{T}_{M}^{N}$, as the integration of
the full 6D EMT in the region with
$r<\epsilon$: 
\begin{equation}
\hat{T}_M^N \equiv \frac{1}{\sqrt{g}|_\epsilon}
\int_0^{2\pi} d\theta
\int_0^\epsilon dr\,
\sqrt{g} L T_M^N, \label{Tdef} 
\end{equation}
and $\hat{T}\equiv \hat{T}_M^M$ as its trace. 

Now we can repeat the procedure with the $\theta\theta$ equation.
Performing the corresponding integration and neglecting again the
terms with only longitudinal derivatives we get
\begin{equation}
\frac{2\pi}{\sqrt{g}|_\epsilon}\int_0^\epsilon dr\, \sqrt{g} L R^\theta_\theta
\simeq
2\pi\left(\beta - \frac{\sqrt{g}|_0}{\sqrt{g}|_\epsilon}\right)
\simeq \frac{1}{M_\ast^4\sqrt{g}|_\epsilon} 
\int_0^{2\pi} d\theta
\int_0^\epsilon dr\, \sqrt{g} L
[T^\theta_\theta-\frac{1}{4} T]
\equiv \frac{\hat{T}^\theta_\theta-\frac{1}{4}\hat{T}}{M_\ast^4}, 
\label{matching:thetatheta}
\end{equation}
where we have defined $\beta(x)\equiv L^\prime(x,\epsilon)$. This
equation, together with eq.(\ref{matching:munu}), determine the
exterior space-time geometry associated with a particular
energy-momentum stored in our codimension two defect. It is apparent
that the $\theta \theta$ equation fixes the deficit angle in the
transverse space, while the $\mu\nu$ equations have the clear
interpretation of requiring a non-zero extrinsic curvature at
$r=\epsilon$ unless $\hat{T}_\nu^\mu - \frac{1}{4}\delta^\mu_\nu
\hat{T}=0$. The trace of this quantity is referred to as the Tolman mass
in 4D cosmic string literature (see $e.g.$ \cite{Frolov:1989er}), and
it is zero for a 
pure 
tension brane. For obtaining these equations we have only neglected
the terms that do not involve $r$-derivatives in the integrals.
 
Notice that the $\mu\nu$ matching conditions are very similar to those
of a codimension one brane, but because $L(x,0)=0$ we cannot satisfy
this equation with a finite $K_\mu^\nu$ in the thin brane limit ($M_b
\rightarrow \infty$) in general, in contrast with the codimension one
case that has a well defined thin limit
\cite{Geroch:1987qn,Israel:1966rt}. In fact it is 
instructive to compare the codimension two case with the more familiar
codimension one case in more detail, and point out their similarities
and differences. For a 5D metric like 
\begin{equation}
ds^2 = g_{\mu\nu} (x,r)dx^\mu dx^\nu - dr^2,
\end{equation}
we can write the $\mu\nu$ components of the Ricci tensor as
\begin{equation}
\sqrt{g} R_\nu^\mu = \frac{1}{2}\left[ \sqrt{g}K_\nu^\mu
  \right]^\prime 
+ \sqrt{g}R_\nu^\mu (g),
\end{equation}
where, as before, $K_{\mu \nu}\equiv \partial_r g_{\mu \nu}$ and
$R_{\nu}^{\mu}(g)$ is the Ricci tensor for the 4D metric $g_{\mu
  \nu}$. We can integrate now this equation from $r=-\epsilon$ to
$r=\epsilon$, a region where we assume that some energy-momentum
density is localized. We get then 
\begin{equation}
\frac{1}{\sqrt{g}|_0}\int_{-\epsilon}^\epsilon \sqrt{g}R_\nu^\mu
=\frac{1}{2\sqrt{g}|_0} \left[\sqrt{g}K_\nu^\mu
  \right]_{-\epsilon}^\epsilon +
\frac{1}{\sqrt{g}|_0}\int_{-\epsilon}^\epsilon \sqrt{g}R_\nu^\mu(g) =
\frac{\hat{T}_\nu^\mu -
  \frac{1}{3}\delta_\nu^\mu\hat{T}}{M_*^3}.\label{matching:cod1}  
\end{equation}
$M_*$ is now the 5D fundamental mass, $\hat{T}_M^N$ is again the
integration of the full 5D EMT in the 
$(-\epsilon,\epsilon)$ region, 
\begin{equation}
\hat{T}_M^N \equiv \frac{1}{\sqrt{g}|_0}\int_{-\epsilon}^\epsilon dr\,
\sqrt{g} T_M^N, \label{Tdef5D} 
\end{equation}
and $\hat{T}$ its trace. It is clear in this case that we can take the
limit $\epsilon \rightarrow 0$ keeping $\hat{T}_{\mu\nu}$, $K_\nu^\mu$
and $R_\nu^\mu(g)$ finite if we accept a discontinuity in the extrinsic
curvature (notice that we still have $\sqrt{g}|_{\pm \epsilon}
\rightarrow\sqrt{g}|_0$ in this limit). We obtain in this way the so
called Israel matching conditions \cite{Israel:1966rt} 
\begin{equation}
\frac{1}{2}M_*^3 \left[K_\nu^\mu \right]_{0^-}^{0^+} = \hat{T}_\nu^\mu -
\frac{1}{3}\delta_\nu^\mu\hat{T}, 
\end{equation}
since only the extrinsic curvature contribution survives in the thin
limit. One could think that the analogy between the thick codimension
two brane and the codimension one case is not surprising since, once 
the brane has been given a certain width, the wall defining the brane
is indeed a codimension one object. But there is some information
in our codimension two matching conditions showing that our system is
different from a codimension one brane\footnote{This difference comes ultimately from the boundary conditions we are imposing at $r=0$, eqs.(\ref{boundary:cond:origin}). If we were imposing instead the boundary conditions $L'(x,0)=0$ and $L(x,0)$ different from zero, we would be describing a 4-brane with a compact dimension ($\theta$) and a $Z_2$ symmetry at $r=0$.}. First, we have an extra 
dimensional matching condition, the deficit angle contribution
eq.(\ref{matching:thetatheta}). As we have mentioned, in the case of a
codimension two pure tension brane with
$\hat{T_r^r}=\hat{T_\theta^\theta}=0$, this contribution absorbs
completely the effect of the brane on the background, allowing us to
keep zero extrinsic curvature. This is not the case for the codimension
one brane, since the right hand side of eq.(\ref{matching:cod1}) does
not vanish for a pure tension brane. This difference is what makes
codimension two braneworlds attractive as a possible solution to the
cosmological constant problem. Also, in the codimension two case we
cannot take the thin limit because $\sqrt{G}=L\sqrt{g}$ vanishes
at $r=0$, and then the left hand side of eq.(\ref{matching:munu}) goes
to zero as $\epsilon \rightarrow 0$ (even allowing for discontinuities
in the first derivatives of the metric) if we insist on keeping
$K_\mu^\nu$ finite. It is then natural to expect 
that, keeping $\epsilon$ finite but small, in a
codimension two brane-world situation 
the extrinsic curvature contribution is still the main contribution to the
integral (\ref{matching:munu}). Thus, we can safely neglect the
integration of the terms involving only $\mu$-derivatives in most
situations, most notably if we are interested in late cosmological
times for which 4D 
curvatures are extremely small.

Another interesting feature of our matching conditions is that
we can now use them in the $\mu r$ equation evaluated at $r=\epsilon$
to obtain an energy-momentum conservation equation for the brane EMT 
\begin{equation}
M_\ast^4 L G_{\mu r}=\frac{1}{2\pi}\left( \nabla^ \nu \hat{T}_{\mu\nu}
+\nabla_\mu \hat{T}^r_r \right)
+ 
\frac{\hat{T}^r_r
+\hat{T}^\theta_\theta}{2\pi} \frac{\partial_\mu M_b}{M_b} -  M_\ast^4
\partial_\mu\left(\frac{ 
\sqrt{g}|_0}{\sqrt{g}|_\epsilon}\right) 
= \frac{T^{bulk}_{\mu r}|_\epsilon}{M_b}.
\label{EMTconservation:general}
\end{equation}
The $\mu r$ component
of the bulk EMT 
determines the flow of energy-momentum from the brane
into the bulk. However, even when that term is zero, there can be an
exchange of energy-momentum between the longitudinal and the
transverse directions (on the brane) correlated with a possible 
space-time dependence of the brane width that could be interpreted by 4D observers as apparent violations of the conservation of energy-momentum.

Up to now we have
performed an integration of the 6D Einstein equations in the
space-time region filled by the brane in order to obtain the matching
conditions. We have identified a set of terms in the equations that
can be integrated in a model independent way, and we have seen that
these terms are indeed the dominant ones provided the $\mu$-dependence
of the solution is 
small. In this way we have 
obtained eqs.(\ref{matching:munu},\ref{matching:thetatheta}) that
relate the first (transverse) 
derivatives of the metric just outside the brane with
the integrated brane EMT. We are taking $g_{\mu\nu}|_\epsilon$ as our
``induced metric'' for the defect, since we are evaluating most
functions in $r=\epsilon$ when dealing with the matching. In fact we
see that all the functions appearing in the matching are evaluated at
the brane boundary except for the ratio
$\sqrt{g}|_0/\sqrt{g}|_\epsilon$ appearing in the $\theta \theta$
matching condition, eq.(\ref{matching:thetatheta}) and in the EMT
conservation equation, eq.(\ref{EMTconservation:general}). In order to
avoid making reference to functions evaluated $inside$ the brane when
dealing with the matching we will consider that this ratio can be
approximated by one in this equation. For this we just need that  
\begin{equation}
\left| \hat{T}_r^r + \hat{T}_\theta^\theta\right| <<
\left|\hat{T}_\mu^\mu\right|. \label{condition} 
\end{equation}
This is because we can put a bound on the difference
$\sqrt{g}|_\epsilon - \sqrt{g}|_0$ as  
\begin{equation}
\sqrt{g}|_\epsilon - \sqrt{g}|_0 \leq \sqrt{g}|_0 \frac{K|_{max}\epsilon}{2}
\sim  \sqrt{g}|_0 \frac{K|_\epsilon}{2 M_b}, 
\end{equation}
where $K|_{max}$ is the maximum value of the function for
$r\leq \epsilon$ and we have used $\partial_r \sqrt{g} =
\sqrt{g}K/2$. Using now the matching conditions, 
eqs.(\ref{matching:munu},\ref{matching:thetatheta}), we arrive at the
mentioned requirement, 
eq.(\ref{condition}). When this condition holds, we could speak of a
quasi-pure-tension-brane, and the extrinsic curvature (times the brane
width) is negligible with respect to the deficit angle. This allows us
to approximate the ratio of the metric determinants in
eq.(\ref{matching:thetatheta}) by one. We quote here the actual 
matching conditions again, assuming such condition holds 
\begin{eqnarray}
2\pi(1-\beta) & \simeq &
\frac{1}{M_*^4}\left(\frac{1}{4}\hat{T}-\hat{T}^\theta_\theta \right)
\simeq   \frac{1}{4 M_*^4}\hat{T}_\mu^\mu, \label{matching2}\\ 
K_\mu^\nu|_\epsilon 
 & \simeq & \frac{M_b}{\pi M_*^4} \left[\hat{T}_\mu^\nu -
\frac{1}{4} \hat{T}
\delta_\mu^\nu \right].
\label{matching1} 
\end{eqnarray} 

We would like to emphasize the generality of these matching
conditions. They apply to \textit{any} codimension two brane, provided
we can neglect the longitudinal $\mu$-derivatives when compared with
the transverse $r$-derivatives in the solution $inside$ the brane, and
our condition eq.(\ref{condition}) holds. 
For instance, it is now straightforward to interpret several known
solutions with 
naked codimension two singularities as being sourced by a codimension
two object with certain energy-momentum. Consider as an example metrics
that near $r=0$ can be approximated as 
\begin{eqnarray}
g_{\mu\nu} &\simeq& \kappa_1 r^{\alpha_1} \eta_{\mu\nu} + \dots,
\label{sing1}\\ 
L &\simeq& \kappa_2 r^{\alpha_2} + \dots. \label{sing2}
\end{eqnarray}
Our matching conditions then imply
\begin{eqnarray}
1- \frac{\kappa_2 \alpha_2}{M_b^{\alpha_2 -1}} & \simeq  &
\frac{T_0}{2\pi M_*^4},\\ 
\alpha_1 &= &  - \frac{1}{4\pi M_*^4}\left(
\hat{T}_\theta^\theta + \hat{T}_r^r\right). 
\end{eqnarray}
where we have taken $\hat{T}_{\mu}^\nu = T_0\; \delta_\mu^\nu$ and we are
assuming that eq.(\ref{condition}) holds. We see how the required
energy-momentum for the defect sourcing these 
solutions depends on the brane width. The brane width acts as a
cut-off for the curvature, and gets rid of the singularity once one
considers a thick defect: remember that these relations have been
obtained matching a regular geometry at $r=0$ (that implies
eqs.(\ref{boundary:cond:origin})) with the exterior geometry given by
eqs.(\ref{sing1},\ref{sing2}). It is interesting to point out that in
the thin brane limit ($M_b \rightarrow \infty$), the brane tension diverges if 
$\alpha_2 <1$ while it goes to zero if 
$\alpha_2>1$ (in this latter case we can not satisfy our assumption in
the thin limit, eq.(\ref{condition}), and the first matching condition
above would have some corrections). We can recognize the only
parameters with a well defined thin 
brane limit yielding a finite brane energy-momentum ($\alpha_1=0$,
$\alpha_2 =1$) as a purely conical geometry. In particular the
singular solutions of 6D supergravity found in \cite{Gibbons:2003di}, 
can be cast in
the form given by ecs.(\ref{sing1},\ref{sing2}) and interpreted as
being sourced by a codimension 2 defect with an energy momentum
tensor given by the formulae above. 

Also, we can now match an exterior AdS geometry with a regular
geometry on $r=0$ and check what kind of energy-momentum distribution
supports such spacetimes. This kind of exterior geometry has been
obtained as the spacetime produced by the Nielsen-Olesen vortex of the
Abelian Higgs model in 6D\footnote{For the global vortex the exterior
  geometry is $AdS_5\times 
  S^1$ \cite{Moreno}, but this case is not a solution for a pure
  cosmological constant in the bulk.} \cite{Gherghetta:2000qi} which,
interestingly enough, localizes not only gravity but also gauge
interactions at the vortex core~\cite{Giovannini:2002mk} (see
also~\cite{Eto:2004ii} for the non-abelian case). Applying
our matching conditions to an exterior geometry like (AdS$_6$ space
with a compact dimension) 
\begin{equation}
ds^2 = e^{\pm kr}\eta_{\mu\nu} dx^\mu dx^\nu - L_0^2 e^{\pm
  kr}d\theta^2 - dr^2,\label{AdSS1} 
\end{equation}
where $k =\sqrt{-\frac{5\Lambda}{2 M_\ast^4}}$, and $\Lambda$ is the
bulk cosmological constant, we obtain the required brane EMT as 
\begin{eqnarray}
1 \mp \frac{L_0 k}{2} e^{\pm \frac{k}{2M_b}} & \simeq &
\frac{T_0}{2\pi M_*^4}, \label{defang}\\ 
\pm \frac{k}{M_b}& = & - \frac{1}{4\pi M_*^4}\left( \hat{T}_\theta^\theta
+ \hat{T}_r^r\right),\label{extrin} 
\end{eqnarray}
where we have taken again $\hat{T}_{\mu}^\nu = T_0\; \delta_\mu^\nu$ and
we are assuming that eq.(\ref{condition}) holds. Notice that if we
choose the minus sign for $\pm k$, so the volume of the spacetime is
finite, we need $T_0>2\pi M_\ast^4$. This condition does not mean that
the curvatures are in the solution bigger than the fundamental mass,
since we expect that $\hat{T}_{\mu}^{\nu}\sim T_{\mu}^{\nu}\epsilon^2
\sim T_{\mu}^{\nu}/M_b^2$. For low values of $M_b$ we can still have
the 4D brane EMT ($\hat{T}_\mu^\nu$) of the order of $M_\ast^4$ or
bigger while the 6D one ($T_\mu^\nu$) is hierarchically smaller and
the curvatures in the full 6D solution are under control, but one can
not keep curvatures under control in the thin brane limit (see also
the discussion in \cite{Gherghetta:2000qi}). However, if 
we want to restrict ourselves to weakly gravitating branes we should
choose the plus sign, and then the extra dimensional volume is
infinite. This is because $L$ is a $growing$ function at the origin
with a positive $r$-derivative, and then it is ``easier'' to match the
geometry with an exterior $L$ function that also has a positive
derivative (and this chooses the plus sign above). 

Our matching conditions relate brane parameters with metric
deformations in its surroundings, but they do not tell us much 
about the
phenomenological viability of these braneworlds. In an ideal
situation, for a given brane energy-momentum and width, and for a
given bulk EMT we would impose the matching conditions in the bulk
solution (as a perturbation, perhaps, over a known static solution)
and find out the implications for the brane induced metric,
$g_{\mu\nu}(x,\epsilon)$. The relation between the brane induced
metric and its EMT would determine what type of gravity brane 
observers would feel. But this approach is usually very difficult to
implement in practice, since it is very hard to find analytic bulk
solutions (even perturbative ones, see \cite{Vinet:2004bk} for work in this
direction). It is however possible to obtain a good deal of 
information on the curvature of the brane induced metric without
actually solving the bulk equations. The idea, 
that we will elaborate on in more detail in the next section for the
cosmological case, is to evaluate Einstein equations just outside the
brane. In particular, the $rr$ component of the Einstein tensor, 
evaluated at $r=\epsilon$ reads,
\begin{align}
G^r_r|_\epsilon &=
-\frac{1}{2} R(g) + 
\frac{M_b^2}{2\pi M_\ast^4} 
[\hat{T}^r_r+\hat{T}^\theta_\theta]
\nonumber \\
&+ \frac{1}{32\pi^2}
\frac{M_b^2}{M_\ast^8} [ 4 \hat{T}^\rho_\sigma \hat{T}^\sigma_\rho -
  (\hat{T}^\rho_\rho)^2 -2 \hat{T}^\rho_\rho (\hat{T}^r_r +
  \hat{T}^\theta_\theta)   
- (\hat{T}^r_r +\hat{T}^\theta_\theta)(5 \hat{T}^r_r -3
\hat{T}^\theta_\theta)]
=\frac{T^r_r|_\epsilon}{M_\ast^4}, 
\label{4Dcurvature:general}
\end{align}
where we have 
used the matching conditions eqs.(\ref{matching2},\ref{matching1})
and have neglected the term $M_b \nabla^\rho \nabla_\rho
L|_\epsilon$. It is not clear \textit{a priori} that this term is
negligible as compared with the induced metric curvature (the
integrals of both terms have been neglected in the matching
conditions). However it is in general related to 
$\mu-$derivatives of $\beta$ that, through our matching conditions, 
can be argued to be negligible under the assumptions we are
using. Again, we will be more specific about the size of the different
terms when discussing cosmological solutions.
This equation determines the curvature for the induced metric in terms
of the brane EMT and the bulk EMT evaluated at the brane
boundary. 
This equation will become
much more illuminating when particularised to the cosmological case as
we will see in the next section. 

\section{Cosmology on a codimension 2 brane.}

In order to study the cosmological implications of our matching
conditions we particularise the metric ansatz to
\begin{equation}
ds^2= N(t,r)^2 dt^2 - A(t,r)^2 d\vec{x}^2 - dr^2 -L(t,r)^2 d\theta^2,
\end{equation}
where we have taken flat spatial sections in the brane for simplicity
and we set $N(t,0)=1$ by performing a redefinition of
the $t$ coordinate. In this section we will assume that
eq.(\ref{condition}) holds in our system, so we can use the simpler
version of the matching conditions,
eqs.(\ref{matching2},\ref{matching1}), instead of the more general
one, eqs.(\ref{matching:munu},\ref{matching:thetatheta}). 
Remember that, since we are assuming that the
first derivatives are small compared with the brane width we also
have, at the level of approximation we are working,
$N(t,\epsilon)\simeq N(t,0) = 1$, and also $A(t,0)\simeq
A(t,\epsilon)\equiv a(t)$. So the matching conditions,
eqs.(\ref{matching2},\ref{matching1}), take now the form 
\begin{align}
& \frac{A^\prime|_\epsilon}{a} = -\frac{M_b}{8\pi M_\ast^4} [\rho+p
    -p_r -p_\theta],
  \label{Aprime} \\
&N^\prime|_\epsilon = \frac{M_b}{8\pi M_\ast^4}
	[3(\rho+p)+p_r+p_\theta],  
\label{Nprime}\\
&(1-\beta)= \frac{1}{8\pi M_\ast^4} [\rho-3p -p_r+3p_\theta]. \label{beta}
\end{align}
where we have taken a ``cosmological'' brane EMT: $\hat{T}^{M}_N =
diag(\rho,-p,-p,-p,-p_r,-p_\theta)$.
These equations, however, do not yield any information about 
the cosmology one can expect in these models. One constraint on it is
given by the brane EMT conservation,
eq.(\ref{EMTconservation:general}), that reads for our cosmological
set up,
\begin{equation}
\dot{\rho} + 3 \frac{\dot{a}}{a} (\rho+p) -\partial_t p_r
- (p_r +p_\theta)\frac{\partial_t M_b}{M_b} =2\pi 
\frac{T^{bulk}_{0r}|_\epsilon}{M_b}. \label{EMTconservation}
\end{equation}
However, we would like to obtain
the equations that govern the evolution of the scale factor of the
brane, $a(t)$, in terms of the brane EMT. As we
said at the end of the previous section, with
our matching conditions we could, in principle, find the bulk solution
for a given model (as a perturbation, perhaps, over a known static
solution) and figure out the implications of this perturbed solution
for the time dependence of the scale factor. Instead of doing that we
will show that a lot of  
information on the cosmology of codimension two branes can be obtained
by evaluating Einstein equations just outside the brane, following the
same approach as \cite{Binetruy:1999ut} for the codimension one case. 
The matching
conditions tell us what the first 
$r-$derivatives of the metric are at the brane boundary,
eqs.(\ref{Aprime}-\ref{beta}), whereas second $r-$derivatives can only be
found by solving Einstein equations. The crucial point to note is
that, out of the five non-zero (for the cosmological metric)
components of the Einstein tensor,
the $tt$, $xx$ and $\theta \theta$ involve
second $r$-derivatives and therefore allow us to algebraically compute
their values at the brane boundary whereas the $tr$ and the $rr$
components do
not involve second $r-$derivatives but only known first $r-$derivatives
and first and second time derivatives of the metric. The former gives
the brane EMT conservation, eq.(\ref{EMTconservation}) whereas the latter,
evaluated at $r=\epsilon$, reads
\begin{align}
\left.G^r_r \right|_\epsilon&=
3 \left ( \frac{\dot{a}^2}{a^2}+\frac{\ddot{a}}{a}  \right)
-\frac{M_b^2}{2\pi M_\ast^4} (p_r+p_\theta)
\nonumber \\
&+\frac{1}{32\pi^2} \frac{M_b^2}{M_\ast^8} 
\left[3
  (\rho+p)^2+(p_r+p_\theta)[2(\rho-3p)-5p_r+3p_\theta]\right]=
\frac{T^r_r|_\epsilon}{M_\ast^4},\label{GenFried}  
\end{align}
where we have particularised eq.(\ref{4Dcurvature:general}) 
to the cosmological case. 
Here we can be a bit more specific about the size of the term we are
neglecting, which is (we do not explicitly write factors of order one)
\begin{equation}
M_b\, \partial_t^2 L=M_b \int_0^\epsilon dr\, \partial_t^2 L^\prime
\approx M_b \, \epsilon \,\partial_t^2 L^\prime|_\epsilon
\approx \frac{\partial_t^2 \rho}{M_\ast^4}, 
\label{ddotL}
\end{equation}
where in the third equality we have approximated (as a conservative order of
magnitude) $\partial_t^2 L^\prime(r\leq \epsilon) \approx \partial_t^2
L^\prime |_\epsilon$, and in the fourth one we have
used the matching conditions.
We therefore see that, at least at late cosmological times it is
utterly negligible as compared with the terms we are keeping.
Nevertheless it should be noted that this is just an order of
magnitude estimation and one should carefully check this approximation
when dealing with particular models (for which this term could play an
important role in the cosmology of our brane).
The equation we have obtained is a generalised
Friedmann equation that 
incorporates the matching conditions for our general codimension two
brane. Taken together with the energy-momentum conservation equation,
eq.(\ref{EMTconservation}), suffices to determine the evolution of the
scale factor 
and therefore the cosmology. As a first check to this equation we
can note that we recover the expected behaviour in the case of a 
pure tension brane with no $rr$ and $\theta\theta$ components of the
EMT. This case corresponds to $\rho+p=p_r=p_\theta=0$ and we get that the 
expansion of the brane depends solely on the bulk EMT and not on the
brane tension, agreeing with the solutions presented in the literature
for this case \cite{Chen:2000at}. 
As a matter of fact, we could have guessed the
generic form of the modified Friedmann equation based on the well
defined infinitesimal limit for a pure tension brane as something like
\begin{equation}
3\left(\frac{\dot{a}^2}{a^2}+\frac{\ddot{a}}{a}\right)= \mathcal{F}(\rho+p) +
\mathcal{G}(p_r,p_\theta,\rho,p)+\frac{T^r_r|_\epsilon}{M_\ast^4},
\end{equation}
where $\mathcal{F,G}$ are arbitrary functions with the only
restriction that $\mathcal{G}(0,0,\rho,p)=0$ and $\mathcal{F}(0)=0$. The bulk
EMT might also have some implicit dependence 
on $\rho,p$ and $p_{r,\theta}$. A critical point when
considering self-tuning issues is the particular form of the function
$\mathcal{F}$. If it was linear in $\rho+p$ it could represent an
important step towards a self-tuning scenario yielding a cosmology of
the type studied in~\cite{Carroll:2001zy} 
whereas the quadratic dependence we have actually found 
would be in conflict with phenomenology. But this negative conclusion
in the self-tuning issue is of 
course 
a bit premature, and should not be taken too seriously, 
in the sense that we have not included the possible
dependence of the bulk EMT on the brane parameters. 
A realistic model might exist that has self-tuning features hidden in
such dependence. For 
the time being however we will naively assume that the bulk EMT does
not have any implicit $\rho$, $p$ or $p_{r,\theta}$ dependence, and we
have just a cosmological constant in the bulk. 
In that case, it
is possible to 
obtain a realistic cosmology if one is ready to give up self-tuning
considerations. 
In our universe, the term
in this equation proportional to $(\rho +p)^2$ would be extremely
small ($\sim \rho_{matter}^2$), while (assuming $p_r$ and $p_\theta$
are constant) the
terms that go like $p_r+p_\theta$ or $(p_r+p_\theta)(5 p_r-3p_\theta)$ 
would act as a
cosmological constant. One expects then that the term $\propto
(p_r+p_\theta)(\rho -3p)$ would give the dominant time-dependence and
therefore a conventional cosmology 
\begin{equation}
3 \left ( \frac{\dot{a}^2}{a^2}+\frac{\ddot{a}}{a}  \right)
\approx
-\frac{M_b^2}{16\pi^2 M_\ast^8}(p_r+p_\theta) (\rho-3p)
+\frac{M_b^2}{2\pi M_\ast^4} (p_r+p_\theta)
+\frac{\Lambda_r}{M_\ast^4},\label{fried}
\end{equation}
where we have just neglected the terms proportional to $(\rho+p)^2$
and $p_{r,\theta}^2$ and defined $T_r^r|_\epsilon\equiv \Lambda_r$. 
There remains, of
course, the issue of the effective cosmological 
constant that would have to be tuned to zero. We have several
parameters that can be chosen at will in the equation above, 
so one can fine-tune the effective cosmological constant to
a small value by requiring
\begin{equation}
\Lambda_r \simeq - M_b^2 \frac{p_r+p_\theta}{2\pi}\left(1-
\frac{T_0}{2\pi M_\ast^4}\right),\label{finetuning} 
\end{equation}
where we have considered that $\rho \sim -p \sim T_0$. There is 
\textit{a priori}
no reason to expect such a
cancellation, so this scenario does not seem to yield any light on the
cosmological constant problem. But notice that the value of $p_\theta$
and $p_r$ coming just from the integration of the bulk cosmological
constant inside the brane has an order of magnitude that, for a weakly
gravitating brane ($T_0/M_\ast^4 <<1$), already agrees with this
requirement: in case of having no ``brane'' contributions to these
quantities one would expect $p_r,p_\theta \sim -2\pi \Lambda_r \epsilon^2
\sim -2\pi \Lambda_r/M_b^2$.

It is very interesting that such a simple model can yield a realistic
cosmology, but one would expect that in more involved models the bulk
deformation produced by the brane EMT would  
affect $T_r^r$ also. Depending on how $T_r^r$ reacts
under a deformed solution (satisfying our $\rho$- and $p$-dependent
matching conditions), one obtains different cosmologies. We see
that considering a constant $T^r_r = \Lambda_r$, $p_r$ and $p_\theta$  
can yield a
conventional cosmology (with the cosmological constant problem
included), but it would be very interesting to see what happens in
other cases. In the case of having two compact extra dimensions stabilised
with a magnetic flux for instance, the deficit
angle in the transverse space affects the local energy density of the
flux, and therefore $T^r_r|_\epsilon$ in eq.(\ref{GenFried}), and this
could be the dominant effect when determining the
cosmology~\cite{Vinet:2004bk}. 
A deeper
study of these issues is currently under way.
Nonetheless, it is interesting to point out that in the case of having
just a cosmological constant in the bulk, 
the effective Planck mass is 
\begin{equation}
M_{Pl}^2= -8\pi^2\frac{M_\ast^8}{M_b^2}(p_r+p_\theta)^{-1} \simeq
4\pi\frac{M_\ast^8}{\Lambda_r}\left( 1 -
\frac{T_0}{2\pi M_\ast^4}\right),\label{Planck} 
\end{equation}
where we have used eq.(\ref{finetuning}). Any relation with the
extra-dimensional volume is not 
transparent in this equation but let's imagine that we are in a
situation whose exterior metric is approximately described by the
geometry given by eq.(\ref{AdSS1}), simply AdS space with a
compact dimension, and the matching conditions reduce to
eqs.(\ref{defang},\ref{extrin}). Taking the minus sign in
(\ref{AdSS1}), so the extra dimensional volume is finite, we can write 
\begin{equation}
V_2 \simeq 2\pi \int_{0}^{\infty} L_0 e^{-5kr/2} dr =
\frac{4\pi L_0}{5 k}. 
\end{equation}
We can use now the matching conditions and the relation $k =
\sqrt{-\frac{5\Lambda_r}{2M_\ast^4}}$ to get 
\begin{equation}
M_{Pl}^2 = \frac{25}{4}M_\ast^4 V_2.
\end{equation}
We need to chose the negative sign in the metric (\ref{AdSS1}) in
order to get a positive Planck mass squared. If we want to match our brane
with an exterior AdS geometry as in
\cite{Gherghetta:2000qi}, and obtain a realistic cosmology, we need that
$T_0>2\pi M_\ast^4$ (that do not necessarily imply curvatures of order
$\sim M_\ast^2$, as we have previously commented).  
If on the other hand we insist on a weakly gravitating brane,
$T_0 << M_\ast^4$, then our matching conditions show that the appropriate
branch is the plus sign in (\ref{AdSS1}) and so the volume of the
extra dimensions is infinite. Interestingly enough, the effective 4D
Planck mass remains finite although imaginary, a situation identical to
the codimension one case with a negative tension
brane~\cite{Shiromizu:1999wj}.  
 Also, as we will see below,  
the dependence on the different scales with high
exponents makes it very easy to generate a hierarchically large Planck
mass without departing from our approximations. 
One can point out again
several similarities between this possibility and the well known
codimension one braneworlds consisting on a 3-brane moving in a 5D AdS
space \cite{Shiromizu:1999wj,Binetruy:1999ut}. 
In both cases the brane EMT enters in the definition of
the effective Planck mass, and in both cases we need a non-zero
``brane vacuum energy'' (the brane tension in the codimension one
case, or the $p_r + p_\theta$ parameter in our case) in order to find a
realistic cosmology. This is a result of the quadratic dependence of
the generalized Friedman equation on the brane EMT parameters in both
cases. Also, in both cases we need to fine tune the bulk 
cosmological constant against brane parameters in order to obtain a
small effective cosmological constant on the brane. It is also thanks to this fine-tuning that we recover the relation of the effective 4D Planck mass with the higher dimensional one times the extra-dimensional volume, as expected from KK arguments (as in the codimension one case). It would be
interesting to push this analogy further, and we might be able to
interpret our thick codimension two brane as a ``curled up''
codimension one brane, since as we have commented previously, the wall
defining the brane boundary is indeed a codimension one hypersurface. 
 
We can also check now the magnitude of the terms we are neglecting in
the Einstein equations when doing the matching for these solutions. In
order to do so, we consider as an example particular values of the
different parameters, motivated by a ``TeV brane''. We consider a
weakly gravitating brane situation, and in this case the exterior
geometry would not be of the type given by eq.(\ref{AdSS1}), since we
need a positive bulk cosmological constant in order to obtain a
realistic cosmology (see eq.(\ref{Planck})). 
Taking for instance the brane parameters to be of the order
\begin{align}
M_b &\approx \mbox{ TeV}, \\
\rho^{1/4}&\approx  (-p)^{1/4} \approx 10 \mbox{ TeV},\\
(-p_r)^{1/4} &\approx  (-p_\theta)^{1/4} \approx \mbox{100 GeV},
\end{align}
and the fundamental scale
\begin{equation}
M_\ast \approx 1.7 \times 10^3 \mbox{ TeV},
\end{equation}
so that the Planck mass comes in the right size
\[
M_{Pl}\approx 10^{18} \mbox{GeV},
\]
we see that the terms we neglected in the matching are of order
\begin{align}
& \int_0^\epsilon dr \; L R(g)  \sim  \frac{3}{M_b^2}\left (
  \frac{\dot{A}^2}{A^2}+\frac{\ddot{A}}{A}  \right)\sim 
10^{-90},
 \label{parallel}\\ 
& \int_0^\epsilon dr \; \partial_t^2 L \sim \frac{1}{M_*^4
    M_b^2}\partial_t^2 (-\rho + 3p -p_r + 3 p_\theta) \sim
  10^{-165}.
\end{align}
We have assumed that the $rr$ component of the bulk EMT is fine-tuned
to give a small (realistic) Hubble parameter, $\Lambda_r^{1/6} \sim
200$ GeV and for the second equation we have used arguments similar to
the ones leading to eq. (\ref{ddotL}).
It is therefore clear that these terms are indeed smaller than
the ones we have considered, that are of order
\begin{align}
& \int_0^\epsilon dr L'' = 1-\beta \simeq \frac{1}{4M_\ast^4}
  (\rho-3p+p_r 
  -3p_\theta)\sim 10^{-9},\\ 
& \int_0^\epsilon dr (L K)' = \frac{K|_\epsilon}{M_b} \sim
  -\frac{2 (p_r+p_\theta)}{M_*^4}\sim 10^{-17}. \label{extrinsic}
\end{align}
Note that these terms, although much larger than the neglected ones,
still have a hierarchy between themselves, as prescribed by our
requirement, eq.(\ref{condition}). 
As we said  we need the small cosmological constant
fine-tuning so that the contributions given by (\ref{parallel}) are
smaller than the ones coming from the terms like
(\ref{extrinsic}). 
One can understand the hierarchy of the
different terms noticing that the terms we neglected in the matching
are in these solutions proportional to the brane EMT squared or its
time derivatives (assuming that the term linear in $p_r+p_\theta$ in
eq.(\ref{fried}) is cancelled by $\Lambda_r$), while the terms we have
kept are matched with the brane EMT 
linearly. One should, however, check for any particular solution the
level of approximation that using our matching conditions represent,
since it is not guaranteed in general that they constitute a good
approximation. In case they do not, our matching conditions have
corrections that depend on the internal structure of the brane.
A numerical estimation of these terms would
be advisable when using our matching conditions,
although as we have seen one can easily find models with solutions in which
they are utterly negligible, and therefore in these models the use of
our matching conditions is fully justified. 

Before finishing this section we would like to stress that this
particular example (and its associated cosmological constant
fine-tuning) has nothing to do with a possible self-tuning
mechanism. In these models, solutions in which the brane is curved
exist on equal footing with solutions in which the brane is flat. So
the required tuning should be provided by extra considerations, the
most promising being 
supersymmetry (see \cite{Burgess:2004ib} and references there in). 
The crucial point is that a
thick brane, via our matching conditions, allows for a dynamical
deficit angle and therefore the possibility of self-tuning is present
in such a set-up. A more detailed study of self-tuning issues in these
models requires knowledge of the bulk solution in more involved models
and is currently under 
way.

\section{Conclusions, open issues and future prospects}

The goal of this letter was to study the dynamics of the induced
metric on a codimension two braneworld, and in particular we were
interested in its relation with the brane EMT. The solution is
singular in general for an infinitesimally thin brane\footnote{As we
  have previously mentioned, this is not the case when one considers
  Einstein-Gauss-Bonnet gravity on the bulk \cite{Bostock:2003cv}, and
  in fact one 
  can find non-singular solutions even for infinitesimally thin higher
  codimension braneworlds when one uses the general Lanczos-Lovelock
  Lagrangians in higher dimensions \cite{Navarro:2004kh}.}, 
except for the case of a
pure tension brane. Contrary to what might seem, 
this case is not relevant for the study of the so called self-tuning
properties on 
codimension two braneworlds due to the staticity (for the deficit angle)
of the solution as
opposed to the intrinsically dynamical nature of self-tuning. Due to
these reasons 
we had to abandon the
thin brane idealization and consider a brane of finite thickness. Our
first step was to find the matching conditions in the second section,
$i.e.$, the set of equations that relate the brane EMT with the
deformation of the surrounding spacetime it produces. Since we are
dealing with a thick brane that has a singular thin limit, it was not
clear that one should be able to do this without knowing all the precise
small-scale structure of the brane. We have however shown that one can
obtain this set of equations depending only on the integral of the
brane EMT as a good approximation provided certain conditions are
met. These conditions have the interpretation of requiring that the
brane has an energy-momentum lying mainly along the parallel
directions (so it is close to a pure tension brane), with small 4D
curvature.  
We have also seen in this second section that the $\mu r$ component
of the Einstein equations gives, when evaluated just outside the brane
using our matching conditions, the energy-momentum conservation
equation for the brane. Using our matching conditions we have been
able to interpret some singular solutions of 6D supergravity found
recently in \cite{Gibbons:2003di} as being sourced by a codimension
two defect with certain energy-momentum and we have also paid
particular attention to the simple case of having just a cosmological
constant in the bulk. In this case we have matched our brane with an
exterior AdS geometry as in \cite{Gherghetta:2000qi}, obtaining the
required brane properties. 

In the third section, attempting to obtain further information that
allows us to assess the phenomenological viability of these models, we
have specialized our equations to the cosmological case. The matching
conditions relate however the brane EMT with the first derivatives of
the metric with respect to $r$, the orthogonal coordinate, and to
obtain the cosmology we would like to relate the brane EMT with the
parallel (time) derivatives of the metric. Fortunately, this can be
done simply by using the matching conditions in the $rr$ component of
the Einstein equation evaluated just outside the brane, at
$r=\epsilon$. The reason is that this component of the full 6D
Einstein equations is the only one (apart from the $\mu r$) 
that does not involve second
$r-$derivatives, so we can get rid of the first ones using our
matching conditions and we are left with only time derivatives of the
induced metric and brane parameters, obtaining the modified Friedmann
equation we were 
looking for. This procedure is completely analogous to the one
followed in the codimension one case by Binetruy \textit{et al.} in
\cite{Binetruy:1999ut}, the only difference 
being the added complication here of having to consider a thick brane,
since the thin limit is singular for the codimension two case. We have
identified a model that could yield a realistic cosmology, simply
considering a constant value for the orthogonal components of the
brane EMT and fine-tuning the bulk cosmological constant to get a
small effective cosmological constant for the brane metric. Put like
that, this
model does not seem to shed any light on the possible self-tuning
behaviour of codimension two brane worlds, 
one of the original motivations for
considering these class of models. 
It would be interesting in any case
to explore this simple model with just a cosmological constant in the
bulk in more detail, in particular to study which bulk 
geometries are obtained from it when we have a positive cosmological
constant in the bulk, that as we saw, makes compatible a weakly
gravitating brane with a realistic cosmology. In case of having a
negative bulk cosmological constant, the exterior geometry is in some
cases just AdS space with a compact spatial dimension
\cite{Gherghetta:2000qi}, and for a pure tension brane the bulk is simply
a wick rotation and analytical continuation of the AdS-Schwarzschild
geometry (see below), so one might expect that the solution asymptotes
to AdS space in general when we have a
negative cosmological constant in the bulk. 

But in fact the main uncertainty, and source of model dependence in our
cosmological equations is the bulk EMT: our modified Friedmann
equation depends on 
its $rr$ component. One can expect that in particular models the
$\hat{T}_{MN}-$dependent matching conditions for the brane imply
that this bulk EMT component also gets an implicit $\hat{T}_{MN}$
dependence, \textit{i.e.} $T_{r}^{r}|_\epsilon$ would also depend on
$\rho$, $p$, $p_r$ 
and $p_\theta$ 
for the cosmological case. It is certainly conceivable that this
dependence is the dominant one, and one could then obtain different
cosmologies depending on the particular model, or compactification,
one is dealing with. A more careful examination of these issues for
different models will be deferred to a future publication. It is worth
pointing out that the bulk and the 
brane curvatures are independent parameters that in principle have no
reason to be related, even for some given brane parameters. This can
be seen by considering the 6D 
black hole with cosmological constant \cite{Birmingham:1998nr},
substituting in it $t \rightarrow i\theta$ and analytically continuing
the constant $(r,\theta)$ hypersurfaces to a Lorentzian manifold of
curvature $H^2$. Then one can interpret the solution as a pure tension
infinitesimally thin codimension 
two braneworld where $H^2$ is the curvature of the brane induced
metric \cite{Wu:2004gp}. One can see then that the brane curvature, the bulk
cosmological constant (that determines curvature of the
bulk) and the brane tension are independent parameters of the
solution. So these models do not have any \textit{selftuning} behaviour
\textit{per se}, if certain models only admit flat 4D geometries
\cite{Gibbons:2003di}, supersymmetry is to blame, and supergravities
with these properties are also known in 7D \cite{Bergshoeff:1985mr}
(where there are 3 compact dimensions and 4 flat dimensions). The
problem, of course, is to obtain an effective 4D theory with
supersymmetry broken at a high scale without spoiling 4D flatness, and
codimension two branes could help in this \cite{Burgess:2004ib}. It
would be very interesting to impose our 
matching conditions in the solutions of the 6D supergravity that have
been proposed in order to realize the selftuning behaviour
\cite{Gibbons:2003di}, and in 
particular to study the implications of these matching conditions for
the supersymmetry of the background, but as we said before, a closer 
examination of these issues is beyond the scope of the present paper,
and will be deferred for future work. 

\section*{Acknowledgements}

We would like to 
thank C.P. Burgess, R. Gregory, H.M. Lee, J. Moreno and A. Papazoglou
for helpful discussions. 
This work has been partially supported by PPARC and by URA under contract
no. \mbox{DE-AC02-76CH03000} with DOE.

\end{document}